# Effects of Ru substitution for Mn on $La_{0.7}Sr_{0.3}MnO_3$ perovskites

L. M. Wang and Jyh-Iuan Wu

Department of Electrical Engineering, Da-Yeh University, Chang-Hwa 515, Taiwan

Y.-K. Kuo*

Department of Physics, National Dong-Hwa University, Hualien 974, Taiwan

**ABSTRACT**

We report the investigations of crystal structure, electrical resistivity ($\rho$), magnetization ($M$), specific heat ($C_P$), thermal conductivity ($\kappa$), and thermoelectric power ($TEP$) on $La_{0.7}Sr_{0.3}(Mn_{1-x}Ru_x)O_3$ (LSMRO) compounds with $x = 0$ to 0.90. From the analyses of crystal structure and magnetization measurements, it is inferred that Ru should have a mixed valence of $Ru^{3+}$ and $Ru^{4+}$ for LSMRO with low level of Ru substitution, and an additional mixed valence of $Ru^{4+}$ and $Ru^{5+}$ with higher Ru substitution. It is found that all measured physical properties undergo pronounced anomalies due to the ferromagnetic-paramagnetic phase transition, and the observed transport properties of LSMRO can be reasonably understood from the viewpoint of polaronic transport. The Curie temperatures $T_C$ determined from the magnetization measurements are consistently higher than those of the metal-insulator transitions $T_{MI}$ determined from the transport measurements. By replacing Mn with Ru, both $T_C$ and $T_{MI}$ decrease concurrently and the studied materials are driven toward the insulating phase with larger value of $x$. It is also found that the entropy change during the phase transition is reduced with more Ru substitution. These observations indicate that the existence of Ru has the effect of weakening the ferromagnetism and metallicity of the LSMRO perovskites.

## I. INTRODUCTION

The hole-doped mixed-valence perovskite manganites of type $R_{1-x}A_x Mn_{1-y}B_y O_3$ (where $R$ is a trivalent rare earth and $A$ is a divalent alkali earth, and $B$ denotes the transition metals) have attracted considerable attention during the past decade. In these perovskite compounds, the interplay between magnetism, charge ordering, and electronic transport have been studied in detail.[1-4] In particular, the metal-insulator transition near the Curie temperature in this class of materials have been interpreted in terms of the double exchange (DE) model, in which a strong exchange interaction occurs between $Mn^{3+}$ and $Mn^{4+}$ ions through intervening filled oxygen $2p$ states.[5,6] In addition, there are some other mechanisms that have provided valuable insight into the colossal magnetoresistance (CMR) phenomenon in the manganites, such as the antiferromagnetic superexchange, Jahn-Teller effects, orbital and charge ordering.[7-9] The knowledge of the crystal structure and the chemical bonding of these compounds is of capital importance to the understanding of the peculiar magneto-transport properties in these perovskites. It is known that the $A$-site substitution changes primarily the carrier density and affects strongly the $Mn^{3+}$-O-$Mn^{4+}$ angle (lattice distortion), thus transforming the parent compound $R$MnO$_3$ from an insulating antiferromagnet into a metallic ferromagnet.[4] On the other hand, the $B$-site substitution is a direct way of modifying the crucial $Mn^{3+}$-O-$Mn^{4+}$ network. The effect of substitution of $Mn^{3+}$ ions in the $B$-site by trivalent ions such as Fe, Ti, and Sc on ferromagnetism and CMR of these manganites has been studied.[10-12] It was experimentally found that any modification on the exchange interaction causes the pair-breaking effect associated with a drastic reduction in Curie temperature $T_C$. For example, $T_C$ reduced by about 22 K for only 1 at. % Fe substitution for Mn in La$_{0.7}$Sr$_{0.3}$MnO$_3$.[10] Other $3d$ transition metals such as Ti, Co, and Ni have similar substitution effect as Fe.[11] However, it was reported that as high as 30 at. % of Ru can be substituted into the Mn sites in La$_{0.7}$Sr$_{0.3}$MnO$_3$ with no change in the crystal structure and has a weak effect on the reduction in $T_C$.[13] It is argued that Ru has a more delocalized $4f$ orbital with itinerant $t_{2g}$ electrons that facilitates the exchange coupling interaction. That is, Ru could make a magnetic pair with Mn to form the Mn-O-Ru network, thus favoring the DE-mediated transport mechanism. Enhanced magnetic and metal-insulator transition temperature in Ru-doped layered magnanites $La_{1.2}Ca_{1.8}Mn_{2-x}Ru_xO_7$ has been reported.[14] Even though a large number of studies on the Ru-doped manganites have been done, the valence state of Ru, which governs the relationships between the magnetism and transport properties of these compounds, is still a subject of debate. Krishnan and Ju reported the X-ray photoelectron spectrum (XPS) of $La_{0.7}Sr_{0.3}Mn_{0.9}Ru_{0.1}O_3$ in which the $Ru^{3+}/Ru^{4+}$ mixed state was suggested.[15] On the other hand, Sahu and Manoharan presented the X-ray absorption spectra of the Ru $L_{2,3}$ edge in $La_{0.6}Pb_{0.4}Mn_{1-x}Ru_xO_3$ and showed a clear signature of the existence of $Ru^{5+}$ up to $x = 0.2$.[16] However, a recent X-ray absorption spectroscopy measurement has given a clear proof that $Ru^{5+}$ is absent in the Na-doped $La_{1-x}Na_xMn_{1-y}Ru_yO_{3+\delta}$ manganites.[17] According to these XPS results the valence state of Ru in manganites has not yet reached a consensus, and thus the source of ferromagnetic interaction between Ru and Mn ions still remains an open question. Therefore, it is worth while examining this important issue from a different point of view. Up to now, although a large number of investigations focused on the electronic and magnetic properties of the LSMRO compounds,[13, 15-19] only a few studies regarding thermal properties or higher level of Ru substitution ($x \geq 0.4$) were reported in the literature. In this paper, we report a throughout investigation of crystal structure, resistivity, magnetization, specific heat, thermal conductivity, and thermoelectric power of $La_{0.7}Sr_{0.3}Mn_{1-x}Ru_xO_3$ with $x = 0$ to 0.90. A discussion of the valence state of Ru based on the results of crystal structure and saturation magnetization is given. Information of such a study makes it possible to determine a full range of physical properties in these Ru-doped magnanites.

## II. EXPERIMENTAL METHOD

Polycrystalline samples of $La_{0.7}Sr_{0.3}Mn_{1-x}Ru_xO_3$ (LSMRO) were synthesized by a conventional solid-state reaction method, using the starting materials of $La_2O_3$, $SrCO_3$, $MnCO_3$, and $RuO_2$ powders. Stoichiometric mixture of powders were ground and reacted at 1100 °C for 24 h in air. The samples were reground, pressed into pellets, and heated again at 1500 °C for 24 h. This procedure was repeated and the samples were finally cooled to room temperature at the rate of 5 °C/min in the last step. The X-ray powder diffraction (XRD) data were collected at room temperature from 20° to 80° with a 2θ step of 0.01° using a diffractometer (Shimazu XRD6000, Japan) with Cu $K_\alpha$ radiation. The resistivities and magnetizations of samples were obtained by a standard dc four-terminal method, and by a superconducting quantum interference device (SQUID) magnetometer, respectively. Relative specific heats ($C_P$) were measured with a high-resolution ac calorimeter, using chopped light as a heat source. Thermal conductivity ($\kappa$) and thermoelectric power (*TEP*) measurements were carried out simultaneously in a close-cycle refrigerator by using a direct heat-pulse technique. The details of the measurement techniques can be found elsewhere.[20]

## III. RESULTS AND DISCUSSION

### A. Crystal structure

Figure 1 shows the typical θ-2θ x-ray diffraction spectra of LSMRO with $x = 0 – 1.0$. As can be seen, the obtained diffraction spectra are consistent with the expected perovskite structure and can be indexed in an orthorhombically distorted structure of space group P2/c. It is noted that for LSMRO with $x > 0.7$, impurity phases such as $SrO_2$ and $La_2O_3$ were detected. The intensity of the strongest impurity peaks is around 5% of the main phase. For $La_{0.7}Sr_{0.3}RuO_3$ ($x = 1$), some diffraction peaks remained unindexed, owing to unidentified impurity phases. Unknown impurity phases have also been observed in $La_{0.5}Sr_{0.5}Mn_{0.5}Ru_{0.5}O_3$ polycrystalline samples.[9] The influence of little impurity phases on the transport properties is thought to be ignorable for LSMRO with $x > 0.7$ due to the insulating impurities of $SrO_2$ and $La_2O_3$. However, the magnetization contributed from the impurity is subtracted. Figures 2(a) and 2(b) show the lattice parameters and the unit cell volume $V_{cell}$ as a function of Ru concentration ($x$) of LSMRO compounds, respectively. It is clearly seen that the lattice parameters $a$ and $b$ increase monotonously with Ru concentration, while the lattice parameter $c$ increases initially for LSMRO for $x < 0.6$ and then decreases and saturates with higher level of substitution ($x \geq 0.6$). The variation of $V_{cell}$ shows similar behavior to that of $c$-axis lattice parameter. The increase of lattice parameters at low levels of Ru substitution has also been observed in Ru-doped $La_{0.67}Ca_{0.33}MnO_3$ samples.[21] The variation of lattice parameters provides us valuable information into a further consideration of the valence state of Ru in these compounds. Since $Ru^{3+}$ (0.68 Å) and $Ru^{4+}$ (0.62 Å) have larger ionic radii compared with that of the $Mn^{3+}$ (0.65 Å) - $Mn^{4+}$ (0.52 Å) pair, it is inferred that Ru should have a mixed valence of $Ru^{3+}$ and $Ru^{4+}$, which could account for the observed increase in the lattice parameters for LSMRO with low level of Ru substitution ($x < 0.6$). In view of the fact that lattice parameters decrease with further Ru substitution ($x \geq 0.6$), it is expected that an additional valence of $Ru^{5+}$ (0.56 Å) might exist in the highly substituted compounds. As previously mentioned, the XPS measurements showed ambiguous results of Ru valence state, we will further discuss this issue in the following section with a different point of view.

### B. Magnetization

Figure 3 shows the temperature dependence of the zero-field-cooled (ZFC) and field-cooled (FC) magnetization for LSMRO. As can be seen, all the LSMRO samples undergo a paramagnetic to ferromagnetic phase transition. It is found that the Curie temperature $T_C$, determined by extrapolating linearly the temperature dependence of magnetic susceptibility $1/\chi$ in the paramagnetic state, decreases with increasing $x$. The variance of Curie temperature $T_C$ versus Ru concentration for $x \leq 0.25$ is tabulated in Table

I. The observed decrease of $T_C$ is in good agreement with that reported in Ref. 15. This magnetic-transition behavior reveals the long-range ferromagnetic ordering up to $x = 0.9$, indicating strong ferromagnetic exchange coupling between Mn and Ru centers. Interestingly, the difference between the ZFC and FC magnetization results is more pronounced with larger Ru substitution level, and the ZFC magnetization shows a sharp decrease at low temperatures. This large thermomagnetic irreversibility has also been observed by others and interpreted as a cluster glass behavior.[12, 21] Such a feature is attributed to a spin freezing below the irreversibility temperature $T^*$, being often observed in inhomogeneous magnetic systems with ferromagnetic grains embedded in a non-ferromagnetic background. On the contrary, the difference between FC and ZFC curves observed in the LSMRO samples becomes less obvious with lower Ru substitution ($x < 0.2$), reflecting the domain-motion character and larger ferromagnetic grains in the samples. The irreversibility temperature $T^*$ decreases with an increase in Ru substitution, as seen in the inset of Fig. 3. These $M(T)$ behaviors are similar to those observed in Ru-doped $La_{0.6}Pb_{0.4}Mn_{1-x}Ru_xO_3$ samples.[16]

Figure 4 shows typical hysteresis loops of the LSMRO compounds measured at 5 K. As can be seen, there is an increase in the coercive field with increasing Ru substitution, implying that the domain wall pinning increases with increasing $x$. Furthermore, the value of saturation magnetization $M_S$ can be approximately obtained from the high-field magnetization. The inset of Fig. 4 shows the magnetic moment $\mu_S$ for the Mn/Ru site estimated from the experimental $M_S$ value as a function of Ru concentration. The obtained $\mu_S$ for the $x = 0$ compound is $3.45\mu_B$, close to the theoretical value of $3.67\mu_B$. It is noted that $\mu_S$ decreases quasilinearly with increase in $x$, but a noticeable slope change is observed at $x = 0.5$. This phenomenon is ascribed to the appearance of different valence states of Ru ions in the low-doping ($Ru^{3+}$) and high-doping regions ($Ru^{5+}$). It has been proposed that $Ru^{3+}$, with a half-filled shell electronic configuration, will result in a local antiferromagnetic coupling with neighboring Mn ions.[17] On the contrary, the introduction of $Ru^{5+}$ will lead to ferromagnetic coupling with Mn spins, which has been proposed for $Sm_{1-x}Ca_xMn_{1-y}Ru_yO_3$ series.[22] With the assumption that a mixed valence of $Ru^{3+}/Ru^{4+}$ in the lower-doping region ($x \leq 0.5$) while an additional mixed valence of $Ru^{4+}/Ru^{5+}$ appears in the high-doping region ($x > 0.5$), we can make a rough estimation for the magnetic moment of these LSMRO samples. The compounds with a chemical formula $La_{0.7}^{3+}Sr_{0.3}^{2+}(Mn_{1-x}Ru_x)_{0.7}^{3+}(Mn_{1-x}Ru_x)_{0.3}^{4+}O_3$ (for $x \leq 0.5$) lead to a ferromagnetic moment of

$$\mu_S (x \leq 0.5) = [4(0.7-0.7x)-3(0.7x)+3(0.3-0.3x)+2(0.3x)]\mu_B = (3.7 - 5.2x)\mu_B, \quad (1)$$

where $\mu_{Mn^{3+}} = 4\mu_B$ ($t_{2g}^3 e_g^1$ state), $\mu_{Mn^{4+}} = 3\mu_B$ ($t_{2g}^3 e_g^0$ state), $\mu_{Ru^{3+}} = 3\mu_B$ ($t_{2g}^5$ state), and $\mu_{Ru^{4+}} = 2\mu_B$ (low-spin $t_{2g}^4$ state) are used. For the higher-doping LSMRO with $x > 0.5$, it can be simply assumed that the increasing Ru concentration has the valence of $Ru^{5+}$. As a result, the chemical formula for the high Ru substituted compounds can be represented by $La_{0.7}^{3+}Sr_{0.3}^{2+}(Mn_{1-x+\frac{3}{7}y}Ru_{0.5})_{0.7}^{3+}(Mn_{1-x-y}Ru_{0.5})_{0.3}^{4+}Ru_{x-0.5}^{5+}O_3$, where $y$ is the reduced component of $Mn^{4+}$ due to the presence of $Ru^{5+}$. According to the conservation of total valence, it can be obtained that $y = \frac{17}{9}(x-0.5)$, and the ferromagnetic moment for $x > 0.5$ is

$$\mu_S (x > 0.5) = (\frac{13}{6} - \frac{6.4}{3}x)\mu_B, \quad (2)$$

where $\mu_{Ru^{5+}} = 1\mu_B$ (low-spin $t_{2g}^3$ state) parallel to the Mn and $Ru^{4+}$ moments is used. Equations (1) and (2) satisfy the situation that (1) = (2) = 1.1 $\mu_B$ with $x = 0.5$. The calculated $\mu_S$ as a function of $x$ is also shown in the inset of Fig. 4 by dotted lines using Eqs. (1) and (2). As can be seen, the calculated $\mu_S$ is in excellent agreement with the experimental $\mu_S$, indicating the validity of our proposal assumption on the variation of valence states of Ru ions in these LSMRO compounds.

## C. Resistivity

The temperature-dependent resistivities ($\rho$ versus $T$) between 20 to 500 K of LSMRO are shown in Figures 5(a) and 5(b) for $x \geq 0.3$ and $x \leq 0.25$, respectively. In general, we found that an increase in Ru substitution for Mn causes a substantial growth on the electrical resistivity of LSMRO. It is seen that the system exhibits metallic behavior (positive temperature coefficient) for low Ru substitution samples ($x \leq 0.15$), but shows semiconducting characteristics (negative temperature coefficient) for high Ru substitution samples ($x \geq 0.25$). With decreasing temperature, noticeable decrease in $\rho$ marks the occurrence of metal-insulator transitions in the low Ru substitution samples. The metal-insulator transition temperature $T_{MI}$, determined from the maximum of $d\rho/dT$, increases initially on Ru substitution ($x = 0.10$), then decreases upon further substitution of Ru ions, and finally disappears with $x \geq 0.50$ in this series of materials, as the resistivity shows no sign of anomaly but exhibits insulating behavior for $x = 0.5$. In Table I, we list the metal-insulator transition temperature $T_{MI}$ with respect to respective compositions for this series of compounds with $0 \leq x \leq 0.25$. A semiconducting-like characteristics for $0.25 \leq x \leq 0.40$ samples below $T_{MI}$'s is observed, and it can be explained by the charge carrier localization due to doping-induced random magnetic potential. On the other hand, an insulating behavior above $T_{MI}$ for $x \geq 0.2$ is observed, leading to a temperature dependence of $\rho$ which can be described by two approaches. One is the variable-range hopping of electrons in a band of localized states,[23] and the other is the polaron formation due to lattice distortion (polaron hopping model).[24] It is found that the data can be fitted into the variable-range-hopping equation,[23] $\rho = \rho_\infty \exp[(T_0/T)^{1/4}]$, to a reasonably good degree with $T_0 \approx (1.4 - 18) \times 10^6$ K and $\rho_\infty \approx (4 - 370)$ $\mu\Omega$ cm. Even though these fitting parameters are comparable with those obtained in Sc-substituting La$_{0.7}$Ca$_{0.3}$MnO$_3$,[12] the obtained prefactor $\rho_\infty$ is unphysical as discussed by Gayathri *et al.* for Co-substituting La$_{0.7}$Ca$_{0.3}$MnO$_3$.[25] Typical $\rho_\infty$ should be of the order of $\rho_{mott}$, the maximum metallic resistivity, which is seldom below 1 – 10 mΩ cm for these oxides.[26] This makes the variable-range hopping an unreasonable proposition for the electrical transport mechanism of LSMRO.

Alternatively, as shown in the inset of Fig. 5(a), we found that the high-temperature resistivity of LSMRO ($x \geq 0.2$) can be fitted by the nonadiabatic small polaron hopping model,

$$\rho = RT^{\frac{3}{2}} \exp\left(\frac{E_\alpha}{k_B T}\right), \qquad (3)$$

where $R$ is a constant, $E_a$ is the resistivity activation energy, and $k_B$ is the Boltzmann constant. The resistivity activation energy $E_a$ as a function of Ru substitution is shown in the inset of Fig. 5(b). As can be seen, $E_a$ increases rapidly with increase in $x$ for $x \leq 0.4$, but decreases in the higher Ru-doping region ($x \geq 0.5$). The obtained $E_a$ values of 103 – 161 meV are comparable with those of 135 – 160 meV for Co-substituting $La_{0.7}Ca_{0.3}MnO_3$.[25] It is known that the small polarons can arise from the strong lattice-electron interaction originating from the Jahn-Teller distortion.[24] This makes a strong correlation between the lattice constant and $E_a$. As mentioned above, the $V_{cell}$ increases initially with $x$ for lower-doped LSMRO, and then decreases with higher level of Ru substitution, as seen in Fig. 2(b). It is apparent that both $E_a$ and lattice constant exhibit similar variation with respect to Ru concentration, suggesting that the polaronic mechanism should be appropriate for the understanding of electrical transport in these LSMRO compounds.

### D. Heat capacity

The $T$-dependent specific heat ($C_P$ versus $T$) of LSMRO with $x$ between 0 and 0.25 is illustrated in Figure 6. Each curve is offset by 30 (J/mol K) for clarity. Note that ac technique does not give the absolute value of specific heat without detailed knowledge of the power absorbed from the light pulse. The absolute value of the specific heat above 130 K is determined by measuring a powder sample (~200 mg) using a differential scanning calorimetry (DSC). The ac results were corrected for their addendum heat capacities (GE varnish and thermocouple wire) and normalized to the DSC data at 200 K. The overall temperature-dependent specific heat of $La_{0.7}Sr_{0.3}MnO_3$ ($x = 0$) is consistent with that reported by Khlopkin et al.[27] It is found that the specific heat undergoes pronounced peaks due to the ferromagnetic-paramagnetic phase transition in these compounds. The transition temperature $T_{c,Cp}$, taken as the temperature of peak position, decreases with increasing Ru substitution level and is found to be consistently lower than $T_{MI}$ and $T_C$, as shown in Table I. Several other reports also suggested that the transition temperature determined from different physical properties may not be the same in these manganites.[28]

The specific-heat jumps $\Delta C_P$ and entropy change $\Delta S$ near the transitions in LSMRO can be estimated by subtracting a smooth lattice background fitted far away from the transition, drawn as a solid curve in Fig. 6. The estimated $\Delta C_P$ decreases with increasing $x$, as shown in the inset of Fig. 6. The corresponding entropy changes $\Delta S$, evaluated by integrating the area under $\Delta C_P/T$ versus $T$ curves, are also found to decrease with increasing $x$. A summary regarding the characteristics of the specific heat anomalies on LSMRO with $0 \leq x \leq 0.25$ is tabulated in Table I. Note that the entropy change ($\Delta S$=0.34 R) for the unsubstituted sample $La_{0.7}Sr_{0.3}MnO_3$ is about half that of the theoretical value $R\ln2$ for a ferromagnetic-paramagnetic phase transition ($R$ is the ideal gas constant). This discrepancy is ascribed to the imhomogeneity of the sample, or partially canted spins in the ferromagnetic state. A small shoulder is noticed around 365 K in the $La_{0.7}Sr_{0.3}MnO_3$ ($x = 0$) sample with our high-resolution ac calorimeter, again presumably owing to the sample imhomogeneity. Such a feature is commonly seen in LSMO polycrystalline samples; however, it becomes essentially invisible for the Ru substituted samples.

The $T$-dependent specific heat for high level of Ru substituted samples ($x = 0.3$ to $x = 0.6$) is shown in Fig. 7. In this figure each curve is offset by 10 (J/mol K) for clarity. It is clearly seen that the transition temperature continuously decreases with increasing Ru substitution and the pronounced peak in $C_P$ gradually evolves into a broad hump for $x = 0.40$, then a slope change for $x = 0.50$. For the $La_{0.7}Sr_{0.3}Mn_{0.4}Ru_{0.6}O_3$ ($x = 0.6$) sample, the anomalous feature is essentially undetectable in $C_P$, being consistent with the observations by electrical resistivity measurement.

### E. Thermal conductivity

Figure 8 shows the observed thermal conductivity ($\kappa$) for LSMRO with $0 \leq x \leq 0.25$. As can be seen, the $T$-dependent thermal conductivity is almost independent of the value of $x$, except for the anomalous feature near phase transitions. The magnitude of $\kappa$ is found to be between 10 to 40 mW/cm K in the temperature range we investigated, consistent with the values seen by others.[29,30] At low temperatures, $\kappa$ increases with temperature and a maximum appears around 40 K for all studied samples. This is a typical feature for the reduction of thermal scattering in solids at low temperatures. With further increase in temperature, $\kappa$ decreases with temperature due to the enhanced phonon-phonon scattering (Umklapp processes), then gradually saturate at high temperatures. Generally, the total thermal conductivity for ordinary metals and semimetals is a sum of electronic and lattice terms. The electronic thermal conductivity $\kappa_e$ can be evaluated using the Wiedemann-Franz law $\kappa_e \rho / T = L_0$. Here $\rho$ is the dc electric resistivity and the Lorentz number $L_0 = 2.45 \times 10^{-8}$ W$\Omega$K$^{-2}$. From this estimation, it is found that the total thermal conductivity is mainly associated with the lattice phonons rather than the charge carriers, due to the high electric resistivity of these perovskites.

As seen from Fig. 8, all samples show distinct anomalous feature due to the occurrence of the metal-insulator transition. The transition temperatures $T_{c,\kappa}$, determined from the maximum of $d\kappa/dT$, are also listed in Table I for comparison. The inset of Fig. 8 shows a blow-up plot near the phase transitions. Since the thermal conductivity measurements provide valuable information about the various scattering processes of thermal carriers, the present data allows us to probe into the interplay between the lattice, charge, and spin degrees of freedom in these manganites. A common feature for the anomaly at $T_{c,\kappa}$ is that a sharp rise in $\kappa$ is observed when the samples enter the ferromagnetic state. The anomalous part of thermal conductivity can be expressed as a combination of electronic, phonon, and magnetic contributions, i.e., $\Delta\kappa = \Delta\kappa_e + \Delta\kappa_{ph} + \Delta\kappa_{mag}$. The electronic term $\Delta\kappa_e$ could be safely excluded from the expression, since the contribution from charge carriers is insignificant as mentioned before. The anomalous enhancement of thermal conductivity at transition could be attributed to the reduction of phonon-phonon scattering due to the Jahn-Teller distortions, which become delocalized along with the charge carriers.[31] However, Ikebe *et al*. have claimed that the phonon alone is far too small to explain the observed enhancement in $\kappa$ at ferromagnetic transitions.[29] Therefore, the magnetic contribution $\Delta\kappa_{mag}$ must be taken into account. It is argued that a suppression of spin-phonon scattering may result from the strong coupling between the lattice and the spin system through the double exchange interaction.[32] Another peculiar feature of the observed *T*-dependent thermal conductivity is that $\kappa$ increases monotonically with temperature in the paramagnetic state ($T > T_{c,\kappa}$). Such kind of behavior is very unusual, since the high-temperature thermal conductivity of crystalline solids is expected to decrease with temperature according to the well-known formula $\kappa \propto \dfrac{aMC\theta_D}{\gamma^2 T}$, where *a* is the lattice constant, *M* is the mass per atom, *C* is the specific heat, $\theta_D$ is the Debye temperature, and $\gamma$ is the Gruneisen constant with a value ranging from 2 to 3 for solids. These parameters are considered to be a weak function of temperature at high temperatures. As a result, $\kappa$ decreases with temperature through a reciprocal function of *T*. However, neutron scattering measurements in these manganites showed that the Gruneisen constant $\gamma$ is not only much larger (~180) than usual, but also decreases markedly with temperature.[33] Thus, the increase in $\kappa$ in the paramagnetic state could be understood with the scenario of local anharmonic lattice distortions.[34]

### F. Thermoelectric power

The temperature-dependent thermoelectric power (*TEP*) data for LSMRO with $0 \leq x \leq 0.25$ are shown in Figure 9(a). The behavior of the *TEP* in these samples shows small and positive values at low temperatures, a signature of *p*-type conduction and metallic in nature. For all samples, it is seen that the measured *TEP* increases initially with *T*, develops into a broad maximum, then changes sign at higher temperatures. This sign change in *TEP* implies that the electronic state of these samples has been thermally altered from hole-like to electron-like. With further increase in temperature, noticeable anomalous feature, which marks the occurrence of phase transition, is clearly observed. The transition temperatures $T_{c,TEP}$ determined from the *TEP* data are tabulated in Table I. In addition to the systematic decrease in $T_{c,TEP}$ with respect to increasing *x* value, the temperatures with the sign change in *TEP* are found to decrease concurrently with $T_{c,TEP}$. Such an observation implies that the sign change in *TEP* is closely related to the magnetic property in these manganites, and can be attributed to the polaron transport, as discussed previously in resistivity. The spin polarization changes the electronic nature of charge carriers, and thus the thermoelectric transport.[32] Moreover, it is conventionally known that poor conductors usually have a larger *TEP* than that of good conductors and vice versa. Near $T_{c,TEP}$, a rapid increase in the absolute value of *TEP* was observed, corresponding well to the metal-insulator phase transition. For *x* = 0.10, a spike-shaped feature was observed in this particular composition, indicating possible enhancement of double exchange interaction at the low level of Ru substitution for Mn in these manganites. Note that the anomalous feature in $\kappa$ was also more pronounced for the *x* = 0.10 sample (see inset of Fig. 9(a)), supporting our previous argument.

In the metallic region, the linear variation of *TEP* is often discussed using the well-known Mott formula

$$S_e = \frac{\pi^2 k_B^2}{3e} T \left( \frac{1}{\sigma(E)} \frac{\partial \sigma(E)}{\partial E} \right)_{E=E_F}, \qquad (4)$$

by assuming a one-band model with an energy-independent relaxation time, where *e* is the elementary charge with a negative sign, and $\sigma(E)$ is the electrical conductivity. In the insulating region, the *TEP* is given by the classical expression

$$S = \pm \frac{k_B}{e} \left( \frac{E_S}{k_B T} + A \right), \qquad (5)$$

where $E_S$ is the activation energy for the thermoelectric power, and *A* is a constant of order unity. In Figure 9(b), we show a plot of *TEP* versus $1/T$ of these manganites and the solid lines represent the fits of measured data to Eq. (5) in the insulating phase. It is seen that the high-temperature *TEP* data can be satisfactorily described using such a thermal-activated model, and the extracted activation energy $E_S$ decreases with increasing *x* (see Table I). Such kind of $E_S$ dependence on *x* has also been found in the partially Cr substitution for the Mn sites in $La_{0.67}Ca_{0.33}MnO_3$.[34] It is worth mentioning that the activation energy $E_a$ in the paramagnetic state is previously estimated to be of the order of 0.1 eV from electrical resistivity data. If Eq. (5) is valid, the *TEP* for these samples should be of the order of 1000 μV/K, which is two orders of magnitude larger than what we observed. With the fact that the thermoelectric power of this class of materials in the insulating phase is unexpectedly small, suggesting a comparable size of electron-pockets and hole-pockets in their energy band and these manganites are nearly compensated. Furthermore, the formation of polarons may also be essential to the thermoelectric transport in these materials.[35]

## IV.  CONCLUSIONS

In conclusion, we have reported detailed investigations of crystal structure, transport

(electrical resistivity, thermal conductivity, and thermoelectric power) and thermodynamic (magnetization and specific heat) properties on $La_{0.7}Sr_{0.3}(Mn_{1-x}Ru_x)O_3$. Based on the crystalline analyses and the magnetic-moment measurements, it is inferred that Ru should have a mixed valence of $Ru^{3+}$ and $Ru^{4+}$ with lower level of Ru substitution ($x < 0.5$), and an additional mixed valence of $Ru^{4+}$ and $Ru^{5+}$ appears in LSMRO with higher level of Ru substitution ($x \geq 0.5$). It is found that all measured physical properties undergo pronounced anomalous features due to the ferromagnetic-paramagnetic phase transition in these compounds. However, the Curie temperatures $T_C$ determined from the magnetization measurement are consistently higher than that of the metal-insulator transitions $T_{MI}$ determined from the transport measurements. By replacing Mn with Ru, both $T_C$ and $T_{MI}$ decrease concurrently and the studied materials are driven toward the insulating phase with larger $x$. The transport properties of LSMRO can be reasonably understood in the viewpoint of polaronic transport. It is also found that the entropy change during the phase transition is reduced with increase in Ru substitution. These observations indicate that the existence of Ru has the effect of weakening the ferromagnetism and metallicity on the LSMRO perovskite. The present study constitutes, by far, the most throughout physical measurements on full range of $La_{0.7}Sr_{0.3}(Mn_{1-x}Ru_x)O_3$ manganites, and provides a significant understanding to the valence state of Ru ions and transport mechanism for these compounds.

## ACKNOWLEDGEMENTS

This work was supported by the National Science Council, Taiwan, Republic of China under contract nos. NSC 94-2112-M-212-001 (L.M.W.) and NSC 94-2112-M-259-012 (Y.K.K.).

*Corresponding author. Email address: ykkuo@mail.ndhu.edu.tw

Table I. Parameters on transport and thermodynamic properties for LSMRO.

| Sample | $T_C$ (K) | $T_{MI}$ (K) | $T_{c,Cp}$ (K) | $\Delta S$ (R) | $T_{c,\kappa}$ (K) | $T_{c,TEP}$ (K) | $E_S$ (meV) |
|---|---|---|---|---|---|---|---|
| $x = 0.00$ | 377 | 362.6 | 360.5 | 0.3376 | 367.1 | 367.1 | 33 |
| $x = 0.10$ | 369 | 362.7 | 358.5 | 0.3153 | 363.4 | 363.4 | 28 |
| $x = 0.15$ | 365 | 344.3 | 340.7 | 0.2993 | 349.0 | 349.0 | 8.1 |
| $x = 0.20$ | 359 | 337.4 | 330.4 | 0.2503 | 337.4 | 337.4 | 6.7 |
| $x = 0.25$ | 328 | 317.7 | 309.7 | 0.1870 | 314.3 | 314.3 | 0.83 |

**FIGURE CAPTIONS**

FIG. 1. Typical θ-2θ X-ray diffraction spectra of LSMRO with $x = 0 - 1.0$. The arrows indicate the impurity phases of $SrO_2$ and $La_2O_3$.

FIG. 2. (a) Lattice parameters and (b) unit cell volume $V_{cell}$ as a function of Ru concentration ($x$) of LSMRO compounds.

FIG. 3. Temperature dependences of the zero-field-cooled and field-cooled magnetization for LSMRO. The inset shows the irreversibility temperature $T^*$ as a function of Ru concentration.

FIG. 4. Hysteresis loops of the LSMRO measured at 5 K. The inset shows the magnetic moment $\mu_s$ for Mn/Ru site as a function of Ru concentration. The dashed lines indicate the $\mu_s$ calculated by Eqs. (1) and (2) for both $x \leq 0.5$ and $x > 0.5$ as a function of $x$.

FIG. 5. Temperature-dependent resistivities between 20 to 500 K of LSMRO with (a) $x \geq 0.3$ and (b) $x \leq 0.25$. Inset of (a): the $\rho/T^{1.5}$ versus $1/T$ plots for LSMRO with $x = 0.2$, 0.5, and 0.7. The solid lines are the fitting of Eq. (3). Inset of (b): resistivity activation energy $E_a$ as a function of $x$. The dashed lines are for guiding the eyes.

FIG. 6. Temperature dependence of specific heats for LSMRO with $0 \leq x \leq 0.25$. Each curve is offset by 30 J/mol K for clarity. The inset shows the specific-heat jumps with backgrounds subtracted.

FIG. 7. Temperature dependence of specific heats for LSMRO with $0.30 \leq x \leq 0.60$. Each curve is offset by 10 J/mol K for clarity.

FIG. 8. Temperature-dependent thermal conductivities for LSMRO with $0 \leq x \leq 0.25$. The inset shows the blow-up plot near phase transitions.

FIG. 9. (a) Thermoelectric power as a function of temperature for LSMRO with $0 \leq x \leq 0.25$. The inset shows the corresponding data at temperatures around the transition temperature. (b) Thermoelectric power *TEP* versus $1/T$ at high

temperatures. The solid lines represent the fits to Eq. (5) in the paramagnetic state.

Fig. 1

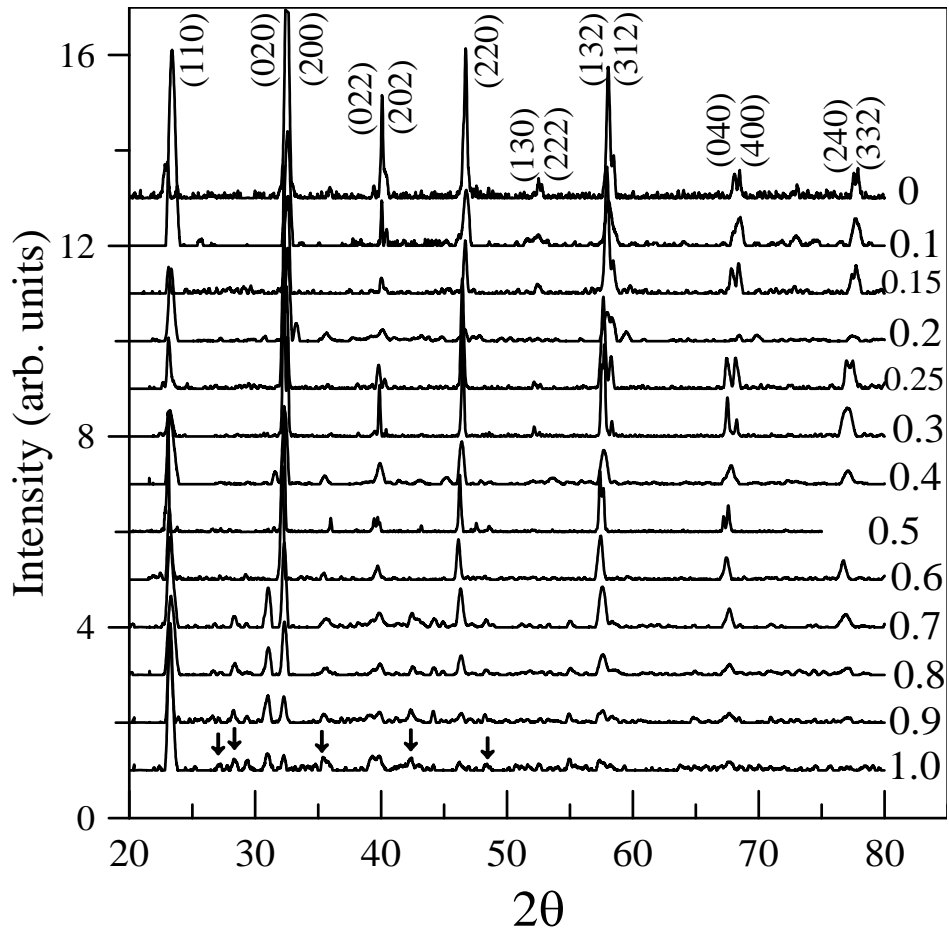

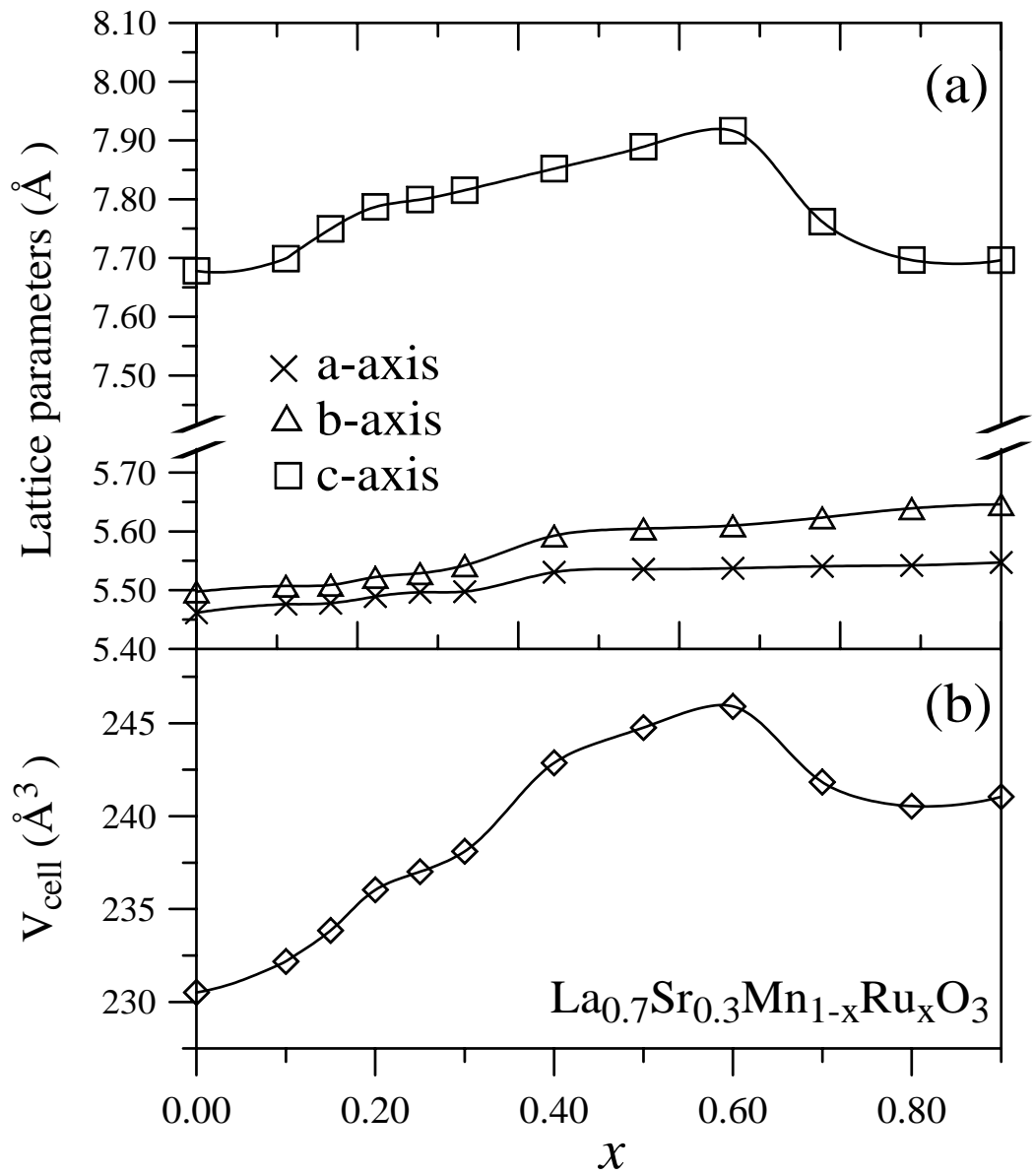

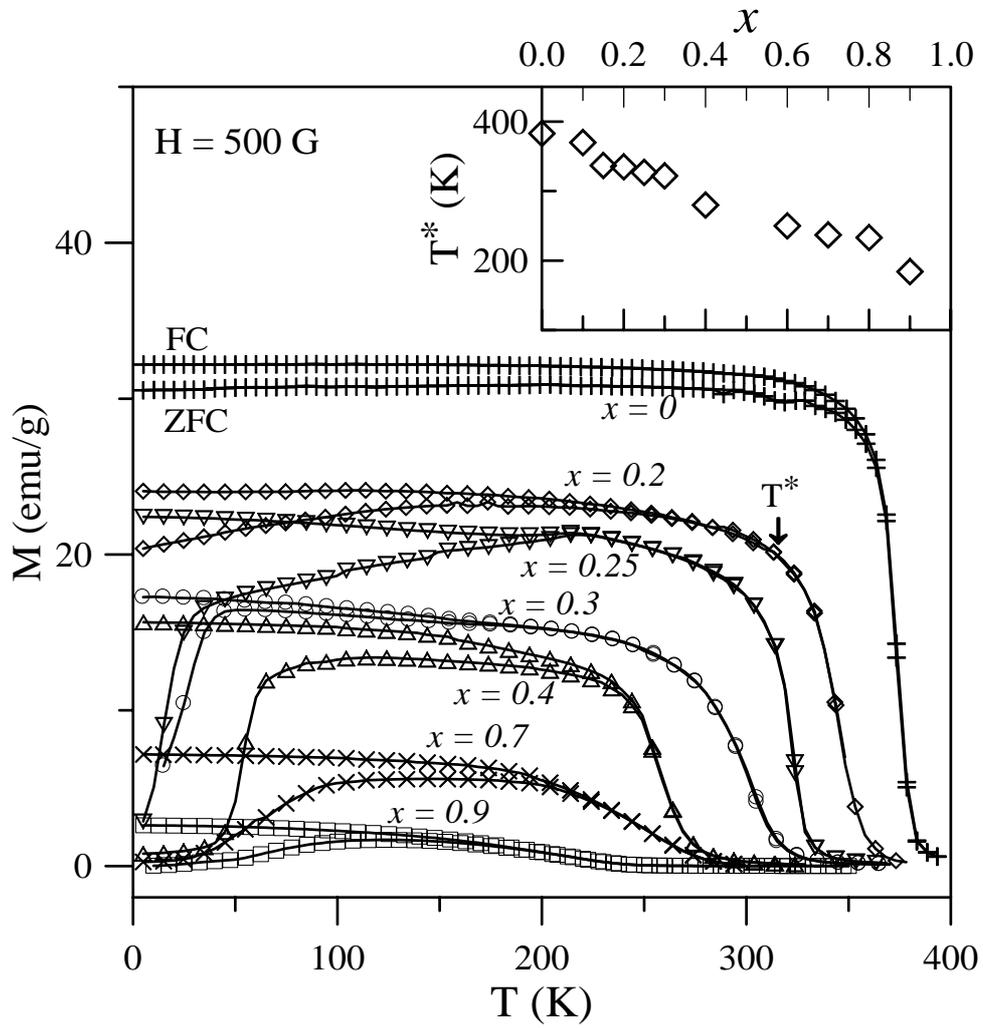

Fig. 4

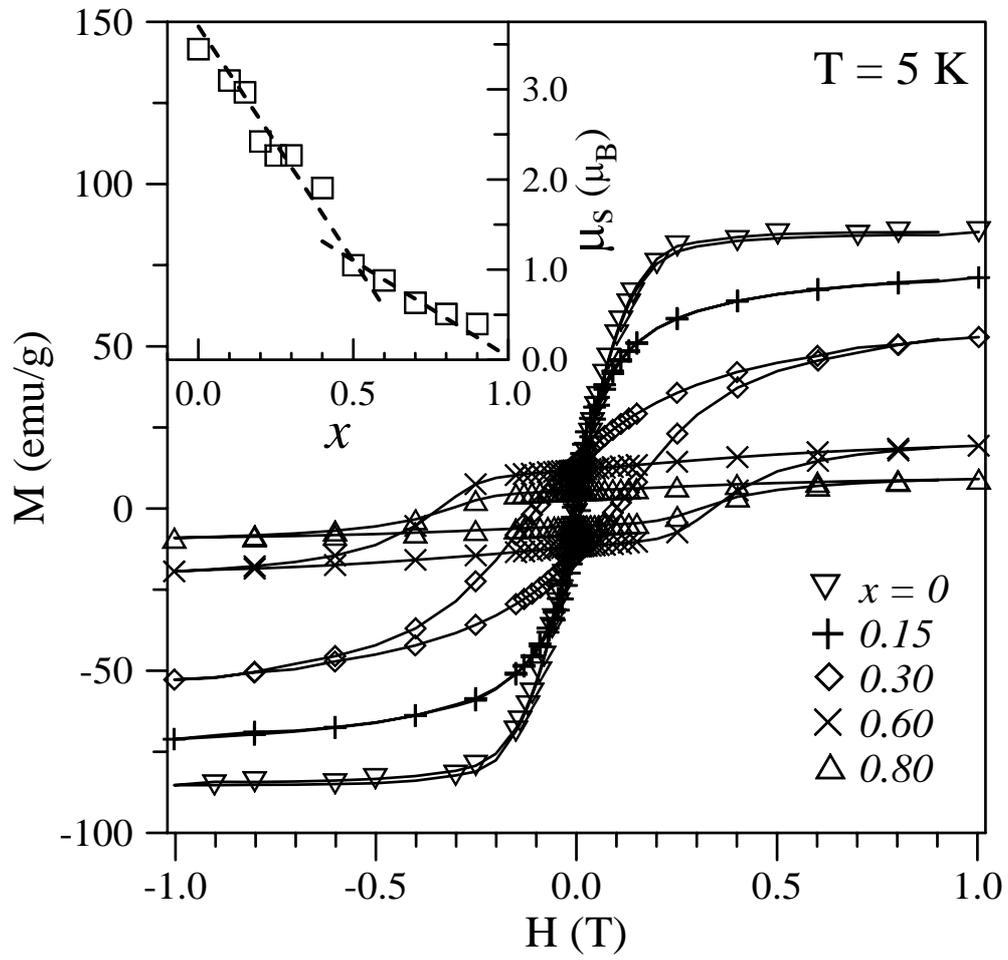

Fig. 5

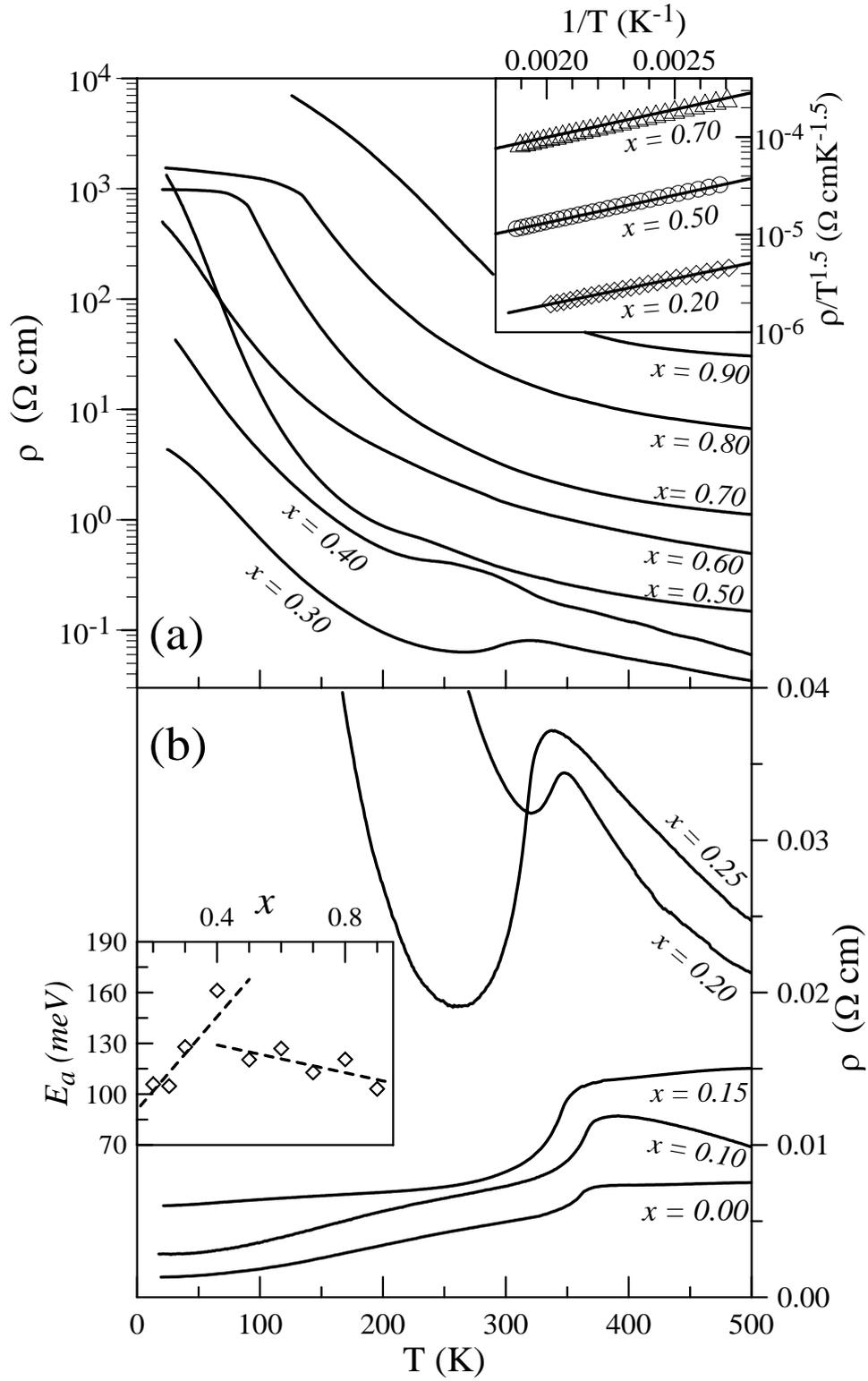

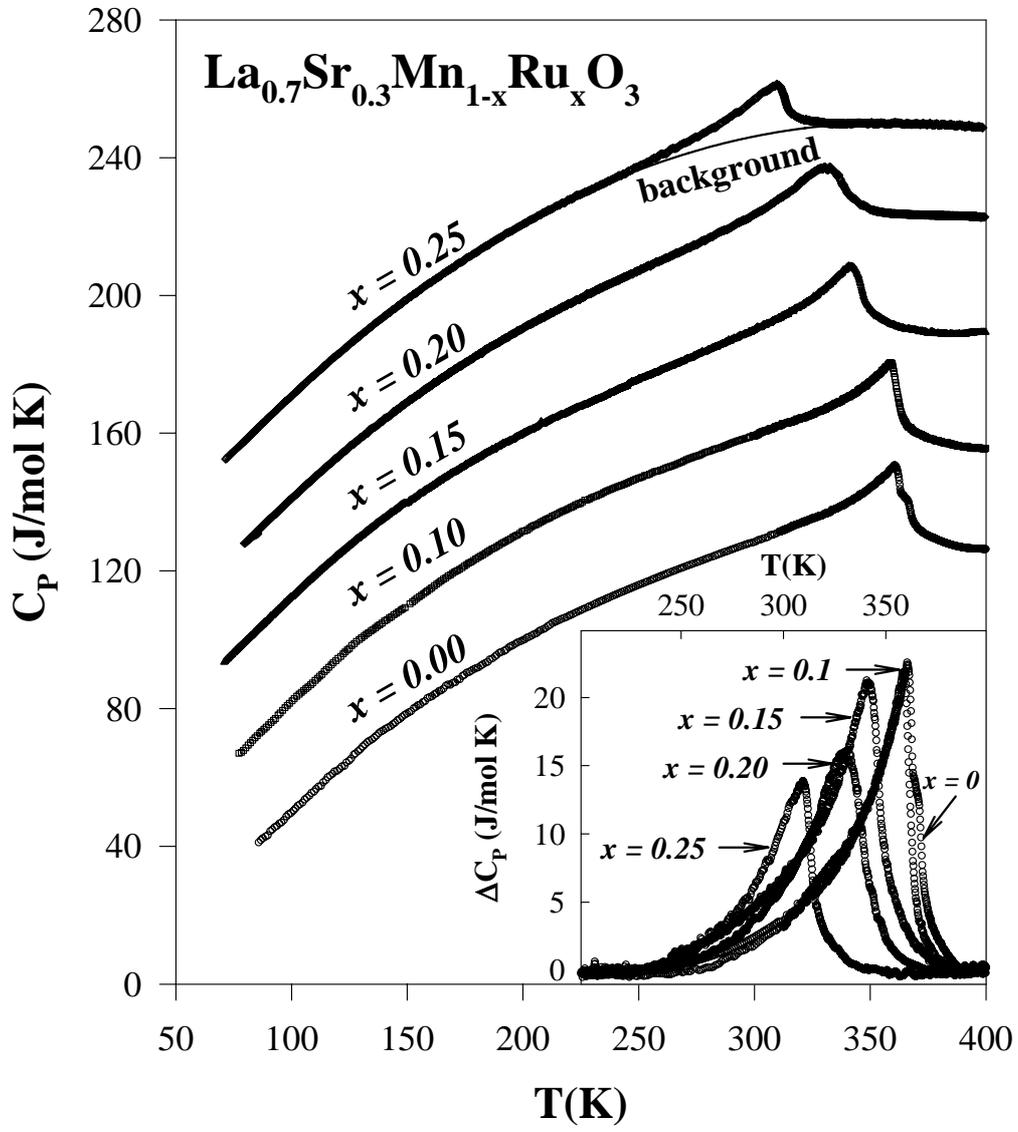

Fig. 7

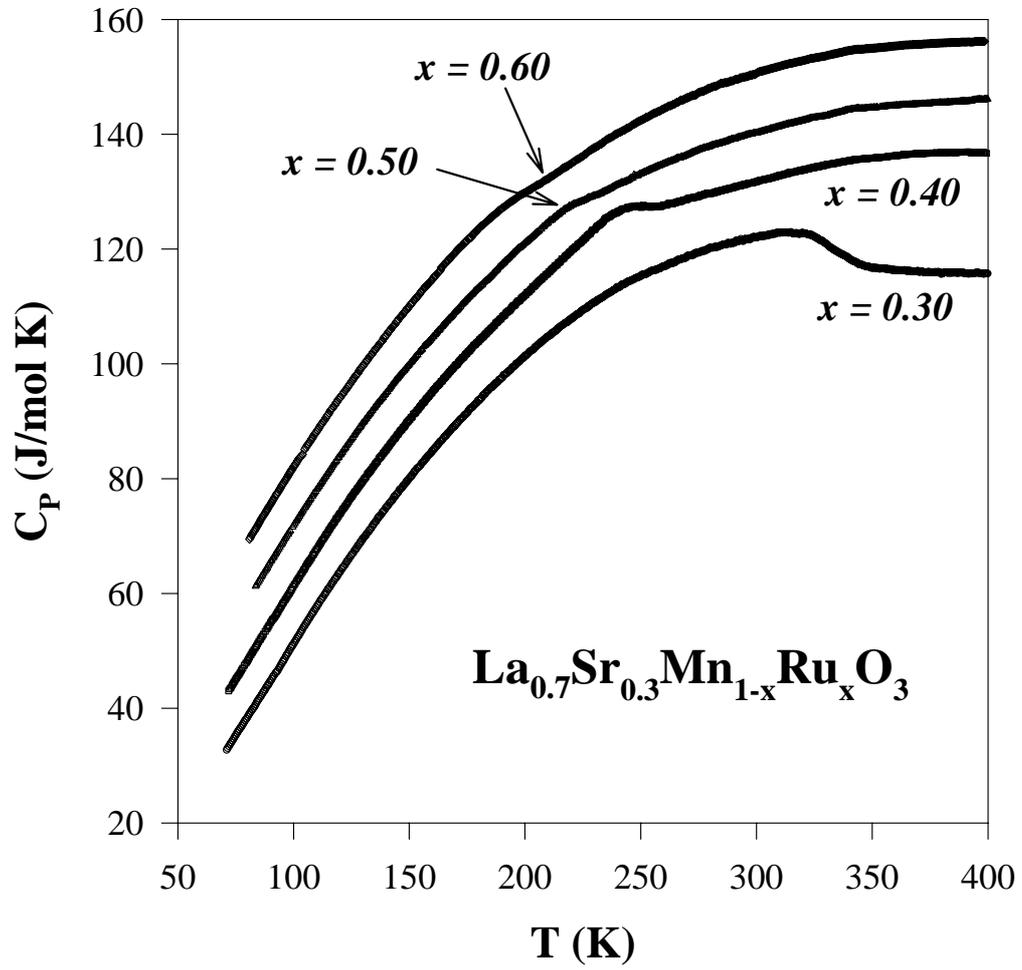

Fig. 8

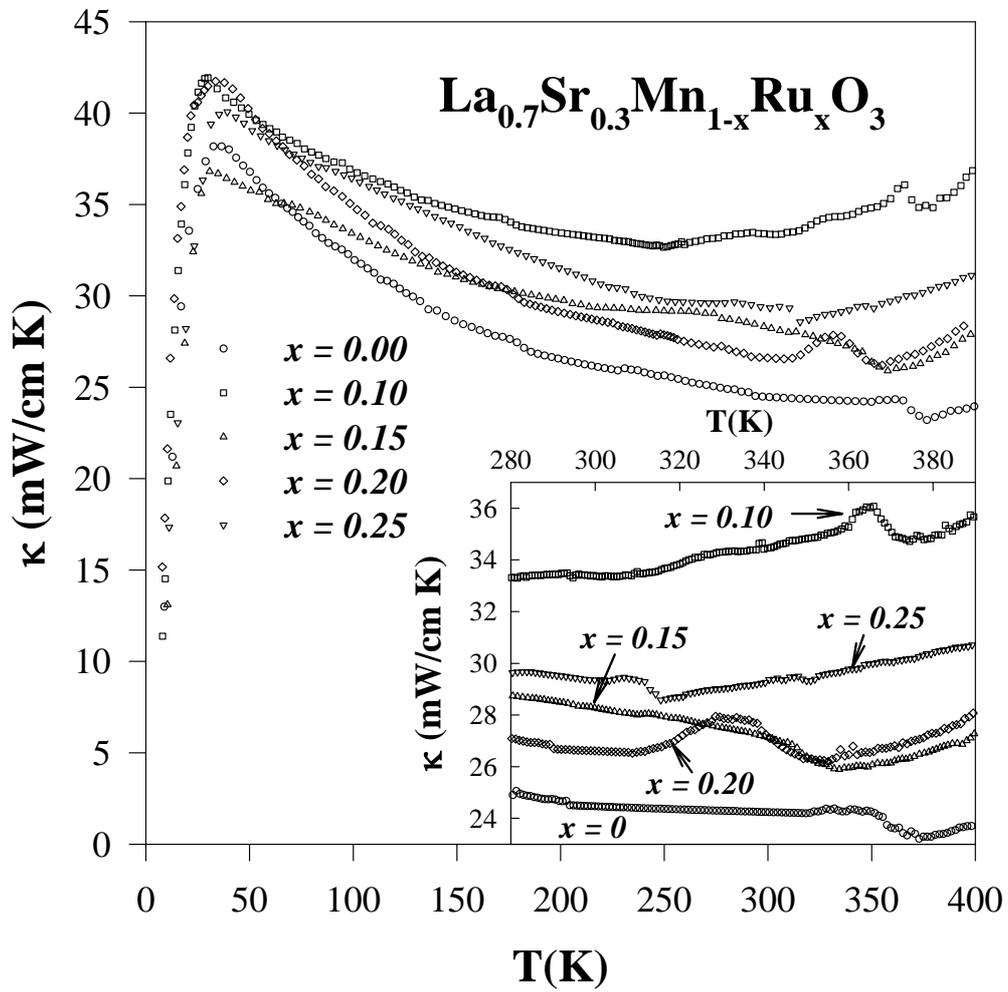

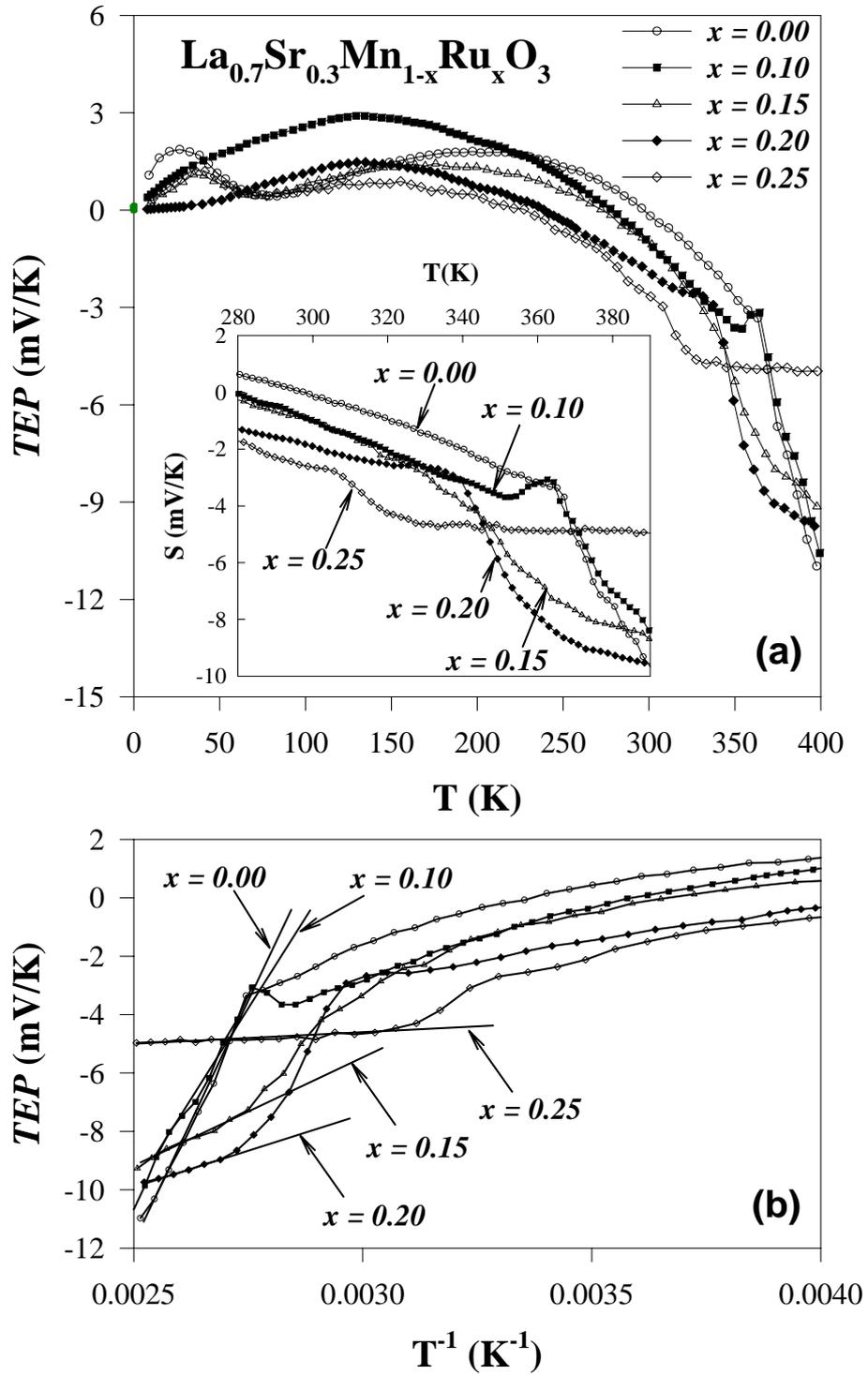